\documentclass[showpacs,preprintnumbers,amsmath,amssymb,twocolumn]{revtex4}
\usepackage{graphicx}
\usepackage{hyperref}

\begin{document}

 \def\[{\left\lbrack}
\def\]{\right\rbrack}
\def\({\left(}
\def\){\right)}
\newcommand{\be}{\begin{equation}}
\newcommand{\ee}{\end{equation}}
\newcommand{\ea}{\end{eqnarray}}
\newcommand{\ba}{\begin{eqnarray}}
\newcommand{\prt}{{\partial}}
\newcommand{\diag}{\mbox{diag}}
\newcommand{\tr}{\mbox{tr}}
\newcommand{\grad}{\ensuremath{\vec{\nabla}}}

 \title{A note on bigravity and dark matter}
\author{M\'aximo Ba\~nados, Andr\'es Gomberoff, Davi C. Rodrigues}
\affiliation{Departamento de F'sica,\\
 P. Universidad Cat\'olica de Chile, Casilla 306, Santiago 22,Chile.}
\author{Constantinos Skordis}
\affiliation{Perimeter Institute for Theoretical Physics, \\
 31 Caroline str. North, Waterloo, Ontario N2L 2Y5,Canada.}

\begin{abstract}
We show that a class of bi-gravity theories contain solutions
describing dark matter. A particular member of this class is also shown to
be equivalent to the Eddington-Born-Infeld gravity, recently proposed as a
candidate for dark matter. Bigravity theories also have cosmological
de Sitter backgrounds and we find solutions interpolating between matter
and acceleration eras.
\end{abstract}

\maketitle

Observations show that most of the energy density in the universe is in the form of dark
matter and dark energy~\cite{Riess:1998cb,Perlmutter:1998np,Spergel:2006hy,Clowe:2006eq,Alcock:2000ph}.
It is therefore of importance to have a simple and natural candidate for these components.

Motivated by Yang-Mills theory, multi-graviton actions are attractive
extensions of general relativity. Consider several metrics $g^a_{\mu\nu}$,
$a=1..n$.  The physical properties and geometric interpretation of such a
theory present great challenges.  It is known that the full (diff)$^{n}$
symmetry cannot be preserved by consistent interactions
\cite{Boulanger:2000rq}. The most general action preserving the full symmetry
group is a sum of decoupled Einstein-Hilbert terms for each metric.

However, interesting theories can be built by breaking the (diff)$^n$
symmetry down to the diagonal subgroup.  For $n=2$, a particular
``bi-gravity" theory with metrics $\{g_{\mu\nu}$,$q_{\mu\nu}\}$ and action
\begin{eqnarray}\label{bigravity}
&& \mbox{\hspace{-.1in}} I = \frac{1}{16\pi G} \int \left\{ \sqrt{-g} (R-2\Lambda) +
\sqrt{-q}(K-2\lambda)  +   \frac{}{}\right. \label{1}\\
 && \left. \frac{1}{\ell^2} \sqrt{-q}\left[ -q^{\alpha\beta}g_{\alpha\beta}
+ \kappa \left((q^{\alpha\beta} g_{\alpha\beta})^2 -
q^{\alpha\beta}g_{\beta\gamma}q^{\gamma\delta}g_{\delta\alpha}\right) \frac{}{}
\right]   \right\} , \nonumber
\end{eqnarray}
has been extensively studied \cite{Isham:1971gm, Isham:1977rj,Damour:2002wu,Damour2002,Arkani-Hamed2003,Blas2006,Berezhiani2007,Berezhiani2008,Deffayet2008}.
Here $\Lambda$ and $\lambda$ are cosmological constants for each sector of the theory. $q^{\mu\nu}$ and $g^{\mu\nu}$ are the inverses of $q_{\mu\nu}$ and $g_{\mu\nu}$ respectively. $K$ is the Ricci scalar for the metric $q_{\mu\nu}$. $\kappa$ is a dimensionless coupling.
A class of bi-measure theories have been considered in~\cite{GK2008} and references therein.

The interaction term above was first proposed in Ref. \cite{Isham:1971gm}
 (see  note \footnote{The map between the constants in (\ref{1}) and those from \cite{Isham:1977rj} reads: $q_{\mu \nu} =  f_{\mu \nu} ~ k_g^2 / k_f^2$, $\frac{1}{\ell^2} = \frac 32 ~ M^2  $, $\kappa = k_g^2 / (6 k_f^2)$.}).
A key feature of this interaction term is that it gives rise to Fierz-Pauli mass terms\cite{fp} for the spin-2 fields.
The above is not the most general mixing term that satisfies this condition. In particular, the density $\sqrt{-q}$  could be replaced by ${(-q)}^u (-g)^{1/2 - u}$ for any real $u$. However, for this theory to
give rise to a dark matter dominated era, we find that, under the assumptions described below, $u=1/2$ is required.

In this short note we point out the following properties of
(\ref{1}). First we prove that for $\kappa=0$ the action (\ref{1}) is
equivalent to the Eddington-Born-Infeld (EBI) theory proposed in
\cite{Banados:2008rm} as a theory for dark matter and dark energy. When one generalizes to the case
$\kappa\neq 0$, it is a natural question to ask whether or not the dark
matter/dark energy interpretation still holds. The answer is in the affirmative.
The metric $q_{\mu\nu}$ can behave both as matter or as dark energy, and
there exist solutions interpolating between them. We  present two
types of de Sitter vacua and study their stability under a certain set of perturbations. One is the well known solution in which the metrics are proportional. In the other case, which has received less attention, the de Sitter line elements of the two metrics are not proportional.
These type of backgrounds were pointed out in different contexts (e.g., in \cite{Damour:2002gp} for the flat case, and in \cite{Blas} in static coordinates). We also display what are the conditions on the couplings that determine whether the Universe evolving from a matter era ends up in the proportional or not proportional vacuum. Finally, we
analyze tensor fluctuations on the de Sitter backgrounds. For the
proportional case the equations can be decoupled and a condition on the
couplings ensuring absence of tachyons is displayed.

The large scale structure of the action (\ref{1}) with $\kappa=0$, called EBI theory, has recently been studied in Ref.~\cite{BFS}. In that article,  it was shown
that the EBI theory has a phase for which the Friedmann background evolution, growth
of inhomogeneities and Cosmic Microwave Background (CMB) angular power spectrum are indistinguishable from those predicted by $\Lambda$CDM. These results  provide extra support for these theories as candidates for dark matter/dark energy. We shall come back to this point at the end.

Theories interpolating between dark matter and dark energy are not new.
Examples are the Chaplygin gas\cite{Kamenshchik:2001cp,Bilic:2001cg,Bento:2002ps}
and the rolling tachyon\cite{Sen:2002nu,Sen:2002in,Gibbons:2002md}.
For the Chaplygin gas, observational consistency of this interpolation has been
challenged in Ref.\cite{Sandvik:2002jz}.

We start by analyzing the relationship between the action (\ref{1}) when $\kappa=0$ and the
EBI theory written in \cite{Banados:2008rm}.  This is straightforward.  In what
follows it is convenient to refer all lengths to $\ell$. We thus define
two new dimensionless couplings $\alpha$ and $\alpha_0$
by
\begin{equation}\label{}
\Lambda = \frac{\alpha_0}{\ell^2}, \ \ \ \ \ \ \lambda= \frac{\alpha}{\ell^2}.
\end{equation}

The first step is to write the action (\ref{1}) in Palatini form with a
connection $C^{\mu}_{\ \alpha\beta}$ associated to the metric $q_{\mu\nu}$.
Varying with respect to $q_{\mu\nu}$, the equations can be algebraically
solved for this field,
\begin{equation}
q_{\mu\nu} =  \frac{1}{\lambda} \left( K_{\mu\nu} - \frac{1}{\ell^2} g_{\mu\nu} \right)
\end{equation}
and therefore the solution can be replaced back in
the action. The resulting action depends on $g_{\mu\nu}$ and the connection
$C^{\mu}_{\ \alpha\beta}$ and is precisely the EBI action,
\begin{eqnarray}\label{max}
  I_{\mbox{\tiny EBI}}[g,C] =   \frac  1{16\pi G} \int  d^4x   \left\{ ~
\sqrt {-g}  \left(R- 2 \Lambda \right)  + \right.\nonumber\\
 \left.  \frac 2 {\alpha \ell^{2}}  ~ \sqrt{ -\det( g_{\mu \nu} - \ell^2 K_{\mu
\nu})}  ~\right\}  .
\end{eqnarray}
Note that the action (\ref{bigravity}), for $\kappa=0$, in its Palatini form, is a {\it parent action} in the sense that one may either eliminate $q_{\mu\nu}$ to obtain EBI, or eliminate the connections to get
(\ref{bigravity}) in terms of the metrics only.

Born-Infeld type actions have appeared repeatedly for many years in many
different contexts. The action (\ref{max}) is particularly close to
the one discussed in \cite{Deser:1998rj} although different in interpretation.  Another class of
class of Born-Infeld theories are tachyonic fields\cite{Sen:2002nu,Sen:2002in}, which are another
candidate for dark matter and dark Energy\cite{Gibbons:2003gb}. This field is described by the
following  effective action,
\begin{equation}\label{tachyon}
I_t = -\int d^4x \; V(\phi) \sqrt{-\det ( g_{\mu\nu} +
\partial_\mu\phi\partial_\nu \phi)} + I_{EH}(g_{\mu\nu}),
\end{equation}
where $ I_{EH}(g_{\mu\nu})$ is the Einstein-Hilbert action including a
cosmological constant and  $V(\phi)$ is an effective potential, which, in
open string theory is\cite{Kutasov:2003er}
\begin{equation} \label{pot}
   V(\phi) = \frac{V_0}{\cosh(a\phi)}.
\end{equation}
Just as the square root in the EBI theory (\ref{max}) can be transmuted into
a standard kinetic term by introducing $q_{\mu\nu}$, a similar manipulation
holds for (\ref{tachyon}). Consider the following action,
\begin{equation} \label{tachyon2}
I_p = V_0 \int \sqrt{-q} \left( -\frac{1}{2}q^{\alpha\beta}\partial_\alpha \phi
\partial_\beta \phi - U(\phi) - \frac 1 2
q^{\alpha\beta}g_{\alpha\beta}\right).
\end{equation}
This action can be seen as Polyakov's version of (\ref{tachyon}), except
that the metric field is $q_{\mu\nu}$. This is seen by varying with respect to
$q^{\alpha\beta}$. We obtain an equation which allows us to algebraically
solve $q_{\alpha\beta}$ in terms of $\phi$,  therefore, we may put it back
in the action (\ref{tachyon2}). We get
\begin{equation} I_p' = \int d^4 x \frac{V_0}{U(\phi )} \sqrt{-\det (
g_{\mu\nu} +
\partial_\mu \phi\partial_\nu \phi)},
\end{equation}
which is precisely the tachyon action (\ref{tachyon}) action when $U(\phi)
=-V_0/V(\phi)$.

Considering the  form of action (\ref{tachyon2}), it is suggestive to add a kinetic
term to the auxiliary metric field $q_{\mu\nu}$.  The obvious choice is to
also add a Einstein Hilbert term with a cosmological constant.  If we do so,
we obtain precisely the first line of (\ref{bigravity}) plus $I_p$ in
(\ref{tachyon2}). This is EBI action in bi-gravity form plus a scalar field
minimally coupled to the metric $q_{\mu\nu}$.
The variation of this action with respect to $q_{\mu\nu}$ gives, again, an
equation that may be solved algebraically for
the $q$-metric. Inserting this back in the action, and redefining $V_0$, we obtain the following
generalization of (\ref{max}),
\begin{eqnarray}\label{max2}
 I &=& \frac  1{16\pi G} \int     \left\{ ~ \sqrt {-g}
\left(R-\frac{2\alpha_0}{\ell^2}\right)  + \right.\\
 && \left.\frac{4}{\ell^{2}(U + 2 \alpha)}  ~ \sqrt{ -\det(  g_{\mu
\nu} - \ell^2 K_{\mu \nu} + \partial_\mu\phi\partial_\nu\phi )}  ~\right\}
.\nonumber
\end{eqnarray}
which becomes EBI when $U=\phi=0$.

Our second goal is to study the cosmological properties of the bi-gravity
system described by (\ref{bigravity}). Assuming that both metrics are
homogeneous and isotropic flat FRW,
\begin{equation}\label{frw}
   ds_g^2 = -dt^2 + a^2 d\vec x^2, \ \ \ \ ds_q^2 = -X^2 dt^2 + Y^2 d\vec
x^2.
\end{equation}
For a given set of couplings $\alpha,\alpha_0,\kappa$ there exists more than one
de Sitter vacua. We distinguish two cases: proportional vacuums (PV) if both metrics are proportional and non-proportional vacuums (NPV).  In the proportional case the functions $a,X,Y$ are given by
\begin{equation}\label{PV}
      a=e^{\frac{H}{\ell}t},\ \ \ \ \ \ \   X^2 = \frac{1- (\alpha_0- 6\kappa) }{1-\alpha}  \ \ \ \ \ \ \ Y = a\, X
\end{equation}
with
\begin{equation}\label{}
 H^2=\frac{1- (\alpha_0- 6\kappa)\alpha
}{3(1-\alpha)} .
\end{equation}
The only condition for the existence of this solution is the positivity
of the constants $H^2$ and $X^2$ above. Note that if $\alpha_0 - 6\kappa =
\alpha^{-1}$, then this vacuum becomes Minkowski. Also note that for
$\alpha$ close enough to $1$, or $\kappa$ sufficiently large, the de Sitter acceleration can be made
arbitrarily large, even if the cosmological constant, $\alpha_0/\ell^2$,
vanishes. Conversely, even for big values of $\alpha_0$ we may fine-tune the couplings in order
to obtain arbitrarily small acceleration. This is an attractive feature in the context of the problem of
the cosmological constant.

A second class of de Sitter vacua with  non-proportional metrics also
exists (NPV).  Let $Y=aXA$, where $A$ is a constant (which is 1 for the
previous case). In this case $a=e^{\frac{H}{l}t}$ where $A$, $X$ and $H$ are
constants determined by the equations:
  \begin{eqnarray}
    \kappa &=& \frac{X^2 A^2}{4} \ \ \ \ \ \  \alpha_0 =  3H^2 + \frac{X^2
A^3}{2} \ \nonumber\\ \alpha &=& -\frac{3}{4X^2A^2} + \frac{3H^2}{X^2} -
\frac{1}{4X^2}
  \end{eqnarray}
  To find the metric parameters $H$, $A$, $X$ one needs to solve a third
order algebraic equation. This means that, in general, we may expect three
different NPV for a given set of couplings. In \cite{Damour:2002gp}, these kind of solutions are also discussed. In that case, however, the cosmological constants are adjusted to have flat backgrounds,
so the mixing term used here, called ${\cal V}_1$ in that reference, gives rise only to the  proportional vacua.

  We may ask now if the above vacua are stable. Consider perturbations of the form
  \begin{eqnarray}\label{}
  a(t) &=& a_{_{0}}(t) + \epsilon a_{_{1}}(t), \nonumber\\
  X(t) &=& X_{_{0}}(t) + \epsilon X_{_{1}}(t), \nonumber\\
  Y(t) &=& Y_{_{0}}(t) + \epsilon Y_{_{1}}(t), \nonumber
  \end{eqnarray}
where the subscript $_{_{0}}$ indicates the background solutions found above. We expand the equations to linear order in $\epsilon$ and look for the conditions on the couplings such that the perturbations do not grow in time. These conditions are best expressed with a picture. In Fig. 1, we show the regions of stability for the PV and the NPV with  $\alpha_0=1/2$ and $\alpha = 0.9$ respectively.

  The vertical axis represent the value of $\alpha$. Note that there are
regions where both solutions are stable. Those are the regions where we have
degenerated vacuum. We do not know, however, if one of them turns out to be
metastable.

\begin{figure}[h]
\begin{center}
 \includegraphics[width=8cm]{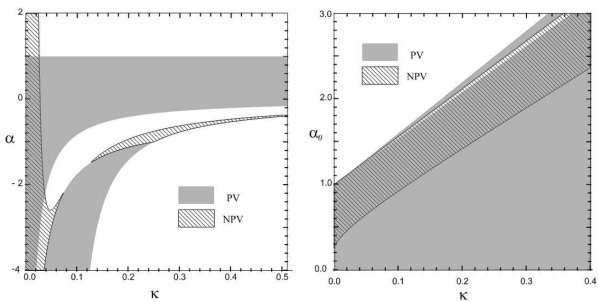}
 \caption{Stability regions for the proportional vacuum  solution (PV) and the non-proportional ones (NPV). In the left the value of
$\alpha_0=0.5$ is fixed. In the right $\alpha=0.9$ is fixed.}
\end{center}
\end{figure}

\vspace{.1in}

 Now we proceed to show, by studying the background cosmological evolution, that this theory encompass a dark matter candidate, even for $\kappa \not= 0$. To this end, the equations of motion coming from (\ref{bigravity}) are evaluated with homogeneous and isotropic metrics. We assume that
both metrics are not singular at the same time \footnote{This ansatz is in close relation to the proposal of having meaningful physics even when $g_{\mu \nu} = 0$ explored in  \cite{Banados:2007qz,Banados:2008rm}.}, which implies that, when the scale factor $a(t)$ [with $a(0) = 0$] is sufficiently close to zero,
\begin{equation}\label{}
 X(t) \approx X_0 + X_1 t, \ \ \ \ \ \ Y(t) \approx Y_0 + Y_1 t,
\end{equation}
where $X_0, Y_0, X_1, Y_1$ are constants. Inserting this into the
equations of motion, with $q$ and $g$ metrics given by (\ref{frw}), one finds $Y_1$ and $X_1$ as a function of $X_0$, $Y_0$, $\alpha$, and $\ell$.
The Friedmann equation for $a(t)$ in the limit of $a(t)$ sufficiently small reads,
\begin{equation}\label{}
3H^2 \approx 8\pi G \rho + \frac{Y_0^3}{  \ell^2 X_0 } \frac{1}{a^3}.
\end{equation}
where $\rho$ is the energy density of other conventional fluids (e.g. radiation or baryons).
Therefore, we see that the q-metric plays the role of an additional dust-like matter, irrespective of the presence of other conventional fluids
like radiation or baryons. Adjusting the constant $ \frac{Y_0^3}{\ell^2 X_0} $ one can have any desired amount of ``dark matter". 
In particular choosing $\frac{Y_0^3}{\ell^2 X_0} = 3.34\times 10^{-7}  w_c Mpc^{-2} $ where 
 $w_c \sim 0.09 - 0.12$ we get the right amount of dark matter as required by cosmological observations.
Note that the $\kappa$ constant has no role at the above regime, and so both the bi-gravity action (\ref{1}) and the EBI model describes the same physics when $a(t) \ll 1$.

Our final task is to study gravitons propagating on the de Sitter vacua. In
the following we use conformal time. We start by perturbing the FRW metrics
as
\begin{equation}
ds^2_g = a^2 \left[ -d\tau^2 + (\gamma_{ij} + h_{ij}) dx^i dx^j\right]
\end{equation}
and
\begin{equation}
ds^2_q =  -a^2 X^2 d\tau^2 + Y^2 (\gamma_{ij} + \chi_{ij}) dx^i dx^j
\end{equation}
where $h_{ij}$ is the tensor mode perturbation of the $g$-metric and
$\chi_{ij}$ the tensor mode perturbation of the $q$-metric. The tensor modes
are transverse and traceless.

From now on we drop the indices on $h_{ij}$ and $\chi_{ij}$ since no
confusion arises.
We find that the field equations for $h$ and $\chi$ are
\begin{eqnarray}
 && \ddot{h} + 2 \frac{\dot{a}}{a} \dot{h} - \grad^2 h =
\nonumber \\
&&
-\frac{2a^2}{\ell^2 \sqrt{-w_0}} \left[ X^2 -2(1-w_0)\kappa\right] (h -
\chi)
\end{eqnarray}
where $w_0 = -\frac{a^2X^2}{Y^2}$,
and
\begin{eqnarray}
 && \ddot{\chi} + \left(3 \frac{\dot{Y}}{Y} - \frac{\dot{a}}{a} -
\frac{\dot{X}}{X}\right)\dot{\chi} + w_0 \grad^2 \chi =
\nonumber \\
&&
-\frac{2a^2w_0}{\ell^2 X^2} \left[ X^2 -2(1-w_0)\kappa\right] (h - \chi)
\end{eqnarray}
respectively.
We now adopt the above equations to the special case of de Sitter vaccua. For
the proportional de Sitter vacuum described above with
$\frac{\dot{X}}{X} = 0$, $\frac{\dot{Y}}{Y} = \frac{\dot{a}}{a}$ and $w_0 =
-1$ these two equations can be collected in matrix from as
\begin{eqnarray}
 \left[\frac{\partial^2}{\partial\tau^2}  + 2 \frac{\dot{a}}{a}
\frac{\partial}{\partial\tau} - \grad^2 + a^2 \mathcal{M}^2
\right]\begin{pmatrix} h \\ \chi\end{pmatrix} =  0
\end{eqnarray}
where the mass matrix $\mathcal{M}^2$ is
\begin{equation}
 \mathcal{M}^2 =  \frac{2}{\ell^2} \left( X^2 -4\kappa\right)
\begin{pmatrix}
1  & -1   \\
- \frac{1}{X^2}   &  \frac{1}{X^2}.
\end{pmatrix}
\end{equation}
One of the eigenvalues is clearly zero while the other one is
\begin{equation}
 m^2 =  \frac{2}{\ell^2} \left(X^2- 4\kappa\right) \left(1 +
\frac{1}{X^2}\right)
\end{equation}
Thus in order for the theory not to contain spin-2 tachyons, we must have $
X^2>4\kappa$ which translates to
\begin{equation}
\frac{  1 -\alpha_0 + 2(1 + 2\alpha ) \kappa }{ 1-\alpha} >  0
\end{equation}
If these conditions are fulfilled, this theory describes a massless and
massive graviton. In particular, they imply stability for the EBI vacuum tensorial modes, which were first studied in Ref.\cite{Rodrigues:2008kv}.

It is important to mention that gravitons are not the only physical
excitations; vector and scalars modes may also propagate. The reason is that
the action has two metrics but only the diagonal subgroup of diffeomorphisms
leaves the action invariant. This means that the scalar and vector modes of
only one of the metrics can be set to zero by a gauge symmetry.  This raises
the issue of the stability of the theory   which should be analyzed along
the lines of Ref.\cite{Higuchi:1987,Deser:2001wx}. It was shown in those references that cosmological
massive gravitons are stable if the mass of the graviton obeys $m^2>2\Lambda/3$.  We  expect similar results to hold in
our case.

To conclude, we have studied in this note  several cosmological aspects of
bigravity actions of the form (\ref{bigravity}). Most importantly we have shown that generically  these actions contain a phase at early times where the second metric behaves as dark matter.  The equations also admit a de-Sitter background which is an attractor if some  conditions on the couplings are fulfilled. This implies a transition between the ``matter" and ``de Sitter" phases.   For the case $\kappa=0$ this transition has been explored in detailed in \cite{BFS} and shown to be problematic at the level of fluctuations and the calculation of CMB spectra.  
However, as shown in \cite{BFS}, one can choose initial conditions and couplings such that the metric $q_{\mu\nu}$ 
is locked into its matter phase up until today. In this case, bigravity predicts a CMB spectrum which is  
indistinguishable from standard particulate dark matter. The calculations for $\kappa\neq 0$ are far more 
complicated and we shall consider them in a separate publication.

\vspace{.1in}
\noindent
{\bf Acknowledgments}

We would like to thank M. Henneaux for a useful discussion and his encourage to consider bigravity actions.
We also thank P. Ferreira for many useful discussions and N. Kaloper for suggestions.
This  work have been supported, in part, by FONDECYT-Chile grants \#1051084, \#1051064, \#1060648, \#7080116 and \#3070008.
Research at the Perimeter Institute is supported in part by NSERC and by the Province of Ontario through MEDT.

\end{document}